\documentclass[runningheads]{llncs}
\usepackage[T1]{fontenc}

\usepackage{graphicx}

\usepackage[hidelinks]{hyperref}

\usepackage{color}

\usepackage{cite}

\usepackage{comment}
\usepackage[title]{appendix}

\usepackage[htt]{hyphenat}

\usepackage[frozencache,cachedir=.,newfloat]{minted}

\usepackage{caption}

\usepackage{tabularx}

\newenvironment{code}{\captionsetup{type=listing}}{}
\SetupFloatingEnvironment{listing}{name=Listing}

\renewcommand{\fcolorbox}[4][]{#4}

\usepackage{orcidlink}

\begin{document}

\title{Spying on the Spy: Security Analysis of Hidden Cameras}

\author{Samuel Herodotou \orcidlink{0000-0002-0382-2902} \and Feng Hao \orcidlink{0000-0002-8664-5074}}

\authorrunning{S. Herodotou et al.}

\institute{Warwick University, Coventry CV4 7AL, United Kingdom
\\
\email{\{samuel.herodotou,feng.hao\}@warwick.ac.uk}}

\maketitle 

\begin{abstract}
Hidden cameras, also called spy cameras, are surveillance tools commonly used to spy on people without their knowledge. Whilst previous studies largely focused on investigating the detection of such a camera and the privacy implications, the security of the camera itself has received limited attention. Compared with ordinary IP cameras, spy cameras are normally sold in bulk at cheap prices and are ubiquitously deployed in hidden places within homes and workplaces. A security compromise of these cameras can have severe consequences. In this paper, we analyse a generic IP camera module, which has been packaged and re-branded for sale by several spy camera vendors. The module is controlled by mobile phone apps available on iOS and Android. By analysing the Android app and the traffic data, we reverse-engineered the security design of the whole system, including the module's Linux OS environment, the file structure, the authentication mechanism, the session management, and the communication with a remote server. Serious vulnerabilities have been identified in every component. Combined together, these vulnerabilities allow an adversary to take complete control of a spy camera from anywhere over the Internet, enabling arbitrary code execution. This is possible even if the camera is behind a firewall. All that an adversary needs to launch an attack is the camera's serial number, which users sometimes unknowingly share in online reviews. We responsibly disclosed our findings to the manufacturer. Whilst the manufacturer acknowledged our work, they showed no intention to fix the problems. Patching or recalling the affected cameras is infeasible due to complexities in the supply chain. However, it is prudent to assume that bad actors have already been exploiting these flaws. We provide details of the identified vulnerabilities in order to raise public awareness, especially on the grave danger of disclosing a spy camera's serial number. 

\end{abstract}

\section{Introduction and Motivation}

Hidden cameras, also known as spy cameras, are digital cameras hidden or disguised as part of common objects, and are generally deployed with the goal to conduct surveillance on people without their knowledge~\cite{yu2022heatdecam}. Although there are legitimate use cases for such cameras (e.g., lawful surveillance on suspects), they can also be misused to spy on people 
unscrupulously. It has been reported that many Airbnbs (1 in 19 in Singapore) have hidden cameras installed, but only 17\% of Airbnb  providers specify where these cameras are located~\cite{airbsns}. Hidden cameras are also frequently installed by parents at homes to monitor the activities of nannies and often the children themselves~\cite{nannycam}. 

The ubiquitous presence of hidden cameras installed in private spaces within homes and workplaces to monitor people without their knowledge clearly raises many privacy concerns. This has motivated many researchers to investigate the detection of such cameras, e.g., via a smartphone's time-of-flight sensors~\cite{sami2021lapd}, a stimulating-and-probing technique~\cite{liu2018detecting}, the analysis of thermal emissions~\cite{yu2022heatdecam}, the RF (radio frequency) signal characteristics~\cite{cunningham2022detection, sindhu2018women}, the Wi-Fi data fluctuations~\cite{salman2022csi, dao2021deepdespy, heo2022there, chaudhary2022demystifying, cheng2018dewicam, lee2022ai, cheng2019detecting}, and the camera's electromagnetic emanations~\cite{liu2023camradar}.  

However, the security of the hidden camera itself has received limited attention. So far, only a few researchers have investigated this subject. Abdalla et al.~show that many cameras use default passwords and the communications are unencrypted~\cite{abdalla2020testing}. Ling et al.~reveal that it is possible to perform an online brute-force attack to uncover the camera's password when the password is only four-digits long~\cite{ling2017end}. They further show that if the MAC address of the camera is known, it is possible to spoof the camera. Biondi et al.~demonstrate that when an attacker is in the same Wi-Fi network as the IP camera, they can eavesdrop on the video data~\cite{biondi2021vulnerability}. Although these studies provide useful insights, their analysis is not systematic, and the identified vulnerabilities tend to have a limited impact. Some of the attacks will not work if the attacker is not in the same network as the camera or if the user changes the default password. 

This paper presents a thorough and systematic analysis of a generic IP camera module, which after repackaging and re-branding, has been built into several best-selling hidden cameras available on Amazon. The camera modules under investigation were purchased at around \$30 each. Some of these hidden cameras are integrated into household objects such as alarm clocks, and are typically sold on Amazon in the range of \$50-120. The camera module is controlled by mobile phone apps that are freely available on iOS and Android. One example is the \emph{LookCam} app, which has over half a million downloads on Google Play alone. However, there are also other apps that work with the same type of module but are branded by different vendors. Security designs for the camera module and the app are not officially published.

By decompiling the \emph{LookCam} Android app and analysing the camera's traffic data, we were able to reverse-engineer the entire security design of the camera system. This includes the Linux operating system (OS) environment on the module, the file structure in the firmware, the authentication mechanism, the session management, and the remote communication with servers in the cloud. Security flaws have been identified in all these areas, and are detailed in Section~\ref{sec:investigation}.

Our contributions are summarised below.
\begin{itemize}
    \item Based on publicly available hardware modules and mobile applications, we have reverse-engineered the security design of a generic hidden camera system. This design does not represent all hidden cameras in the market but is believed to be fairly common among commercial products.  
    \item Based on the reverse-engineered security design, we have identified categorical flaws and presented proof-of-concept attacks accordingly. These flaws allow an adversary to perform remote code execution on a camera from anywhere in the world with the mere knowledge of the camera's serial number.
    \item Based on the findings, we propose mitigation measures and good practices for designing more secure camera systems in the future.
\end{itemize}

\subsubsection{Ethics and responsible disclosure} The camera modules being analysed were purchased and owned by the authors. Proof-of-concept attacks were demonstrated against these devices only without affecting other IP cameras in use. We responsibly disclosed the findings to the manufacturer. Whilst the manufacturer acknowledged our work, they showed no intention to fix the problems, mainly because patching/recalling these modules is infeasible due to complexities in the supply chain. On the other hand, the public needs to be informed of the risk of using hidden cameras, especially since users sometimes share serial numbers of the purchased cameras in online reviews. One CVE (Common Vulnerabilities and Exposures) has already been assigned (\textit{CVE-2023-30400}), and others are also under review at the time of writing. The following sections will detail the vulnerabilities with the manufacturer's name anonymised. 

\section{Hardware and Supply Chain}

\begin{figure}
    \centering
    \includegraphics[width=\linewidth]{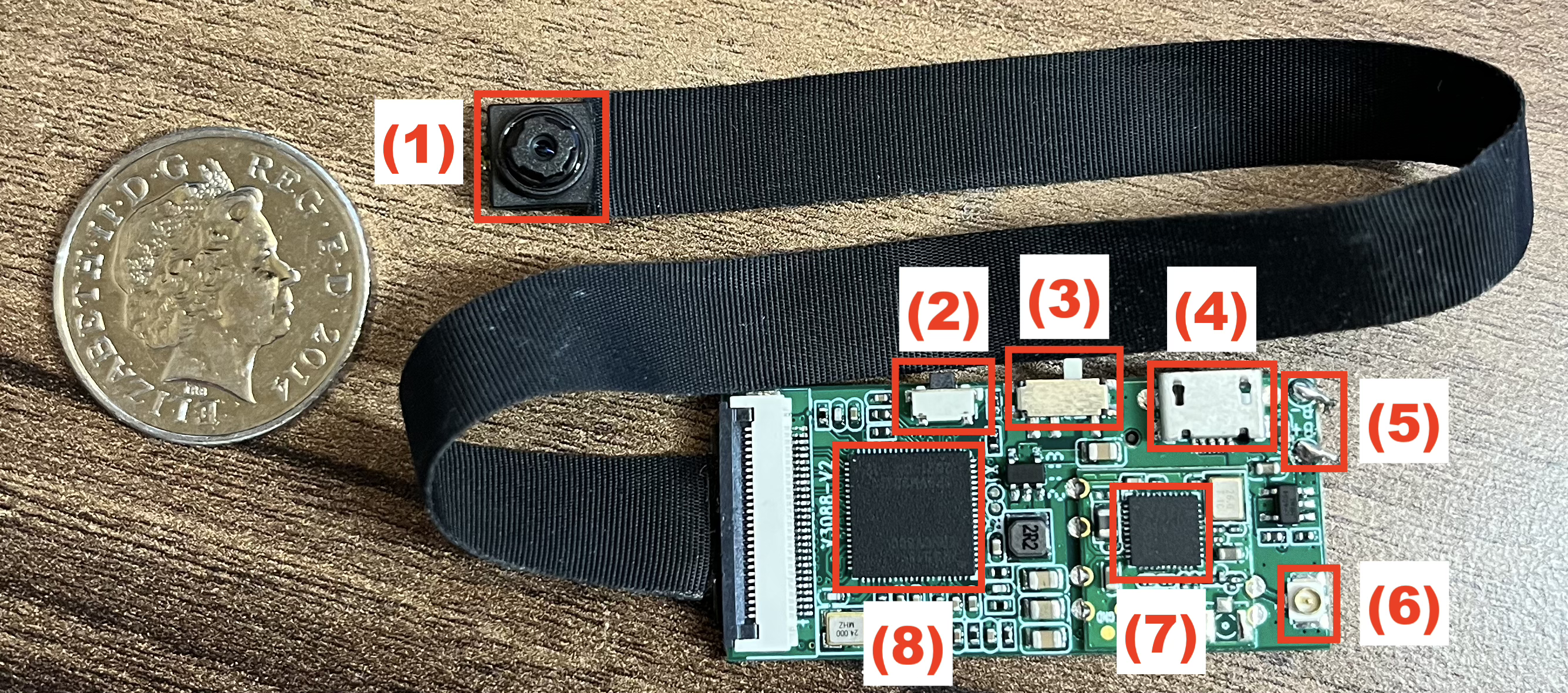}
    \caption{The Camera Module. (1) Micro camera. (2) Reset button. (3) Power switch. (4) Micro-USB port. (5) Power pins (battery). (6) Wi-Fi antenna. (7) Wi-Fi module. (8) Central Processing Unit.}
    \label{fig:camera_module}
\end{figure}

The generic camera module under analysis is a portable, thumb-sized device that can be powered with a battery or micro-USB. It works completely standalone, supporting live video streaming and Wi-Fi connectivity out-of-the-box. Optionally, a Micro SD card can be inserted to enable video recordings. Figure \ref{fig:camera_module} shows a photograph of the camera module.

The device is designed to connect with a companion app, which is developed by vendors under different brands. The app analysed in this investigation is called \textit{LookCam}. Its features include live streaming, remote configuration, and downloading previously recorded footage. 

The modules in question originate from a prominent firm in the electronics industry, referred to hereafter as \emph{the manufacturer}. This manufacturer 
specialises in the production of camera modules and CCTV (closed-circuit television) equipment, and according to 
publicly available information online, exports \$5-10 million worth of product yearly, with their main markets in Europe, America and Asia. 

In terms of the supply chain, this manufacturer acts as the OEM (original equipment manufacturer). The modules are sold in bulk to other vendors, which are then packaged and re-branded. The final products are released to consumers in online stores such as \textit{Amazon}.
After the generic camera modules are sold in bulk, even the manufacturer cannot track where these modules are distributed to third-party sellers at multiple retail levels. The complexities in the supply chain have profound security implications since if there is a security flaw in the generic module, it is virtually impossible to patch or recall the affected products. 

This manufacturer also partners with two other companies in producing the camera modules.
One is a leading integrated circuit manufacturer. They produce the system-on-chip,
which is a core component of the camera module, providing an embedded-Linux operating system and drivers to support an IP camera product.
The other company specialises in providing a peer-to-peer networking system,
which is a software component of the camera module responsible for facilitating remote connections to the cameras. Serious flaws have been discovered in these components as well. According to public information available on the company's website,  
the peer-to-peer networking system has been adopted by over 50 million IoT devices.

\section{Investigation}
\label{sec:investigation}

This section describes the testbed setup, the reverse-engineering process, and the vulnerabilities identified with 
proof-of-concept attacks.

\subsection{Pairing the Device}

To pair a camera with the mobile app, there are multiple approaches. When no network is configured (e.g., if reset to factory settings), the device hosts its own hotspot network which the user can connect to. Once connected, the \textit{LookCam} app can automatically pair by listening for packets sent by the device (which contain its serial number). Alternatively, a user can add a device that is already connected to the internet by supplying its serial number to the app. It is common for these devices to include a sticker or QR code which contains the serial. When connecting via the app, the user will be prompted to enter a password to gain access. All devices are configured with a default password of \texttt{123456}.

\subsection{Testbed Setup}

\begin{figure}
    \centering
    \includegraphics[width=0.9\linewidth]{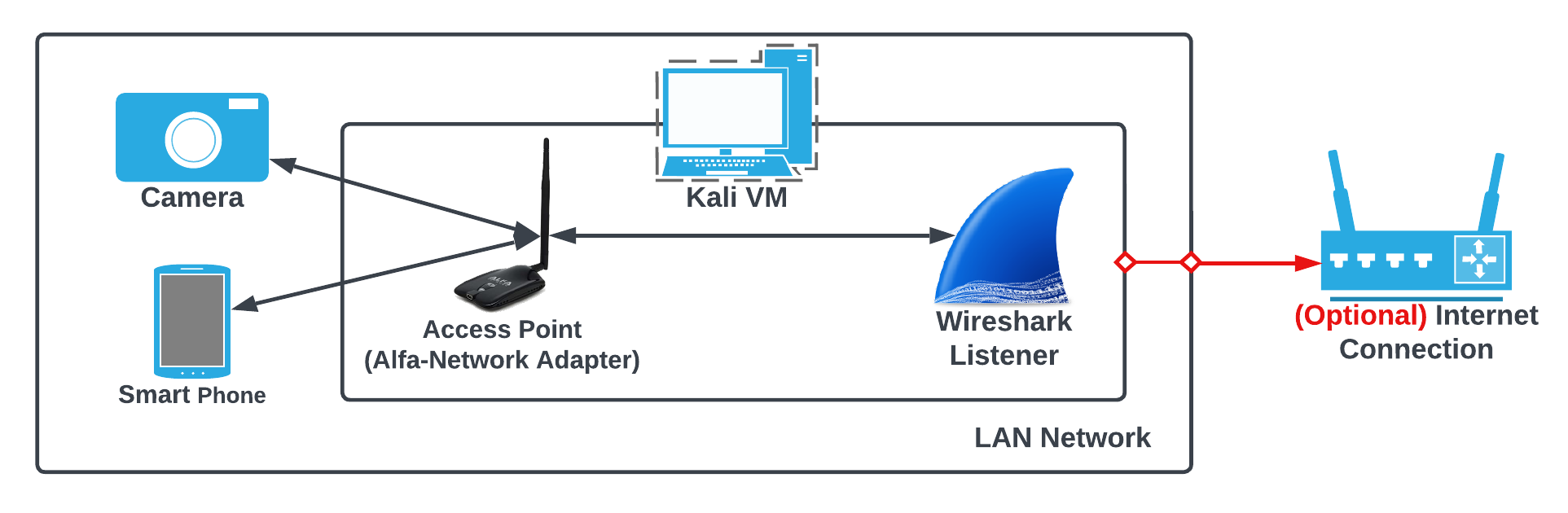}
    \caption{Architecture of the network sandbox used to intercept traffic}
    \label{fig:testbed}
\end{figure}

To facilitate an investigation of the network services running on the device, it is necessary to construct a network sandbox to intercept all relevant communications. This was achieved by connecting an external wireless network adapter (\textit{Alfa-Network AWUS036NHA}) to a \textit{Kali Linux} virtual machine. By using the \textit{hostapd} tool, a custom Wi-Fi hotspot was created with the adaptor. Configuring the camera module to connect to this network would then enable all communications to be intercepted using a packet-sniffing tool such as \textit{Wireshark}. By using an additional network adaptor to create the hotspot, the built-in network adaptor of the Kali machine could be used to bridge an Internet connection to the hotspot, enabling all external traffic to be intercepted (e.g., communications with peer-to-peer servers). See Figure \ref{fig:testbed} for a diagram of the structure. This testbed was set up only for reverse-engineering the security design of the camera system. For attacking the system, the adversary does not need to be in the same Wi-Fi network as the camera; the attack can be launched from anywhere on the Internet.  

\vspace{-0.8cm}
\subsection{Mobile Application Analysis}

Without knowledge of the camera's security design, the reverse-engineering process started with analysing the controlling app, in particular, the \textit{LookCam} Android app that is publicly available in Google Play. Decompiling the \textit{LookCam} Android application with \textit{Jadx} enabled its source code to be analysed. From an initial scan, it was discovered that the core of the networking functionality is implemented within a C-library named \texttt{libPPCS\_API.so}. Investigating this library required disassembly in \textit{Ghidra}, and is discussed later in this paper.

Additionally, a \emph{secret} logging feature was discovered in the \texttt{AboutActivity.java} file. This file controls how users can interact with the `About' page in the app. The code reveals that, if a user holds down the `LookCam' logo for a few seconds, a menu is revealed that allows the user to export a debug log. Other applications were also discovered to include this functionality. This log contains output from all the components of the application, including the C-libraries that communicate directly with the camera. This log output provided vital information on how the phone communicates with the camera module, and revealed a JSON (JavaScript Object Notation) command system in use. Listing \ref{lst:debug_output} provides a portion of the output, revealing the structure of a login command sent to the camera.

\begin{code}
\begin{minted}[breaklines,frame=single]{text}
LookCam[28765:1775458] Connect Success!! SessionID=34
LookCam[28765:1775458] will login with session 34
LookCam[28765:1775519] mediaDataRecThread going...
LookCam[28765:1775458] send json {
    cmd = LoginDev;
    pwd = 123456;
}
\end{minted}
\vspace{-1em}\captionof{listing}{Log output revealing a JSON-style command system in use}\vspace{1em}
\label{lst:debug_output}
\end{code}

\subsection{Unencrypted Communications}

By analysing the network traffic produced during interactions between the camera module and the app, a UDP (User Datagram Protocol) service running on port 32100 was discovered. Monitoring network traffic whilst using the app revealed that the service provides all of the core functionality of the module, from configuration to live-streaming video. This was possible since the protocol transmits all data in plaintext, 
enabling an eavesdropper to read all communications between the camera and the app. This includes 
sensitive information such as login requests (containing the device’s password in plaintext), the contents of configuration commands, and live video footage. 
Once the attacker has intercepted the device’s password, they can gain full access to the camera via the mobile application as if they were a legitimate user. 
However, exploiting this flaw is not easy as it requires the adversary to be a man-in-the-middle (MITM) between the camera and the phone. However, this MITM requirement no longer becomes a constraint when exploiting vulnerabilities in the camera's peer-to-peer and command systems, enabling the camera to be controlled from anywhere on the Internet. These vulnerabilities are discussed in the following sections.

\subsection{Vulnerable Command System}

Many flaws were discovered in the JSON command system, used by the app to interact with the camera. A custom client was developed to mimic the actions of the mobile phone app, which enabled custom JSON payloads to be sent that could exploit potential vulnerabilities in the implementation of the command handlers.

\subsubsection{Bypassing Authentication}
To begin, an analysis of communications between the camera and the app revealed that the camera's password is included in every request made by the app. This is included in plaintext under the \texttt{pwd} field in the JSON body. Not only does this increase the probability of an eavesdropper capturing the device's password, but highlights a lack of session management in use by the system. Listing \ref{lst:json_commands} demonstrates the standard format used by all commands sent by the app.

\begin{code}
\begin{minted}[breaklines,frame=single]{json}
{
    "cmd": "[Command name]",
    "pwd": "[Device password]",
    "...": "...",
}
\end{minted}
\vspace{-1em}\captionof{listing}{JSON structure of commands}\vspace{1em}
\label{lst:json_commands}
\end{code}

When sending an incorrect value for the \texttt{pwd} field, one would expect the camera to reject the command completely. However, using the custom client to send malformed commands with the \texttt{pwd} field omitted revealed that the camera makes no attempt to verify the supplied credentials. This shows that the user's password authentication is performed client-side in the app, and not on the camera. Although the \texttt{LoginDev} command is sent to the camera to verify the supplied password, this command simply verifies the correctness of the password without updating the state of the system or establishing a session. This makes it possible for an attacker to gain full access to the camera without knowing the password by using a custom client, similar to the one developed in this investigation. Alternatively, using dynamic instrumentation tools such as \textit{Frida} makes it possible to disable the code responsible for performing the client-sided check. The \texttt{loginDevice} function within \texttt{LuPPCSSession.java} was successfully hooked and overwritten to bypass this check. This eliminates the need for an adversary to develop a custom app from scratch to bypass the authentication system. Thus, by adding any known serial number to the \textit{LookCam} app with this custom code enabled, an attacker can gain full access to a target camera without being on the same network or being a MITM.

\subsubsection{Reading Configuration Values}
\begin{figure}
    \centering
    \includegraphics[width=\linewidth]{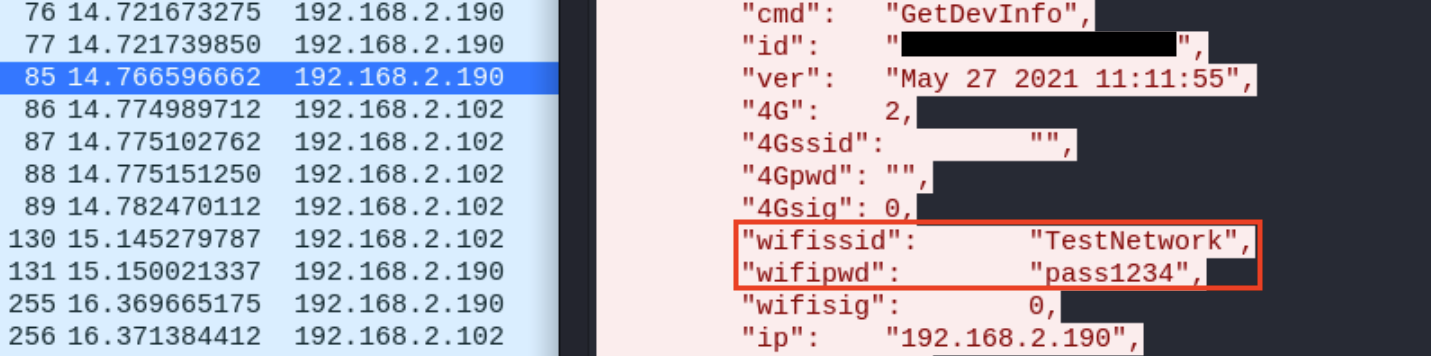}
    \caption{Extracting Wi-Fi credentials in Wireshark}
    \label{fig:dev_info}
\end{figure}

Given that full access can be granted without knowing the password, an attacker no longer needs to perform a man-in-the-middle attack and rely on user-interaction for sensitive information to be obtained. This information can be requested directly, as the device cannot distinguish an attacker from a legitimate user. The \texttt{GetDevInfo} command can be sent, which is then responded with sensitive information such as the user's Wi-Fi credentials, as shown 

in Figure \ref{fig:dev_info}. The transmission of the Wi-Fi credentials to the app appears totally unnecessary, which demonstrates a lack of security-consciousness from the manufacturer in the security design.

\subsubsection{Live Streaming} By imitating the requests the application makes when requesting a live-stream, an attacker can access live-footage from a target camera without the user's interaction. 
Many of these cameras also include microphones, enabling audio to be captured too. 
Even the installer of the camera may not be aware that the spy camera can be spied on by random people on the Internet. This clearly aggravates privacy concerns about these hidden cameras. 

\subsubsection{Arbitrary File Downloading} The camera module offers a file-downloading command to facilitate the remote retrieval of historic footage. 
A vulnerability was discovered in the file-download command handler that enables attackers to download arbitrary files present on the system. See Listing \ref{lst:download_file} for an example file-download request.

\begin{code}
\begin{minted}[breaklines,frame=single]{json}
{
    "cmd": "DownloadFile",
    "pwd": "123456",
    "patch": "/mnt/CYC_DV/20220708@111673.mp4",
    "pos": 0
}
\end{minted}
\vspace{-1em}\captionof{listing}{Example file download request}\vspace{1em}
\label{lst:download_file}
\end{code}

By sending modified requests with a custom client that was designed to mimic the \textit{LookCam} app, it was possible to send \textit{any file path} under the \texttt{patch} parameter. The camera immediately responds with a series of UDP packets containing the contents of the file. No attempt was made by the manufacturer to sandbox the file system or ensure file paths are within the recording directory. This makes it possible to download any file on the device, as long as the path is known. Recalling that this can be performed without the user's password, the scope in which an attacker can extract data is no longer limited by what information the network service is designed to share. For example, the file \texttt{/etc/jffs2/.devpsd} was discovered, which stores the user's password in plaintext.\footnote{In some newer devices, this is stored in \texttt{/etc/config/.devpsd}.} The lack of encryption in this file makes it possible for an attacker to effortlessly obtain this information. This breach of confidentiality could pave the way for further malicious activities, as the password may be reused on other systems.

\subsubsection{Shadow File Extraction} It was possible to download the shadow file located in \texttt{/etc/shadow} using the file-downloading vulnerability. The shadow file is a protected file that stores the password hashes for Linux users.\footnote{We note that these hashes are unrelated to the device password used by the app to authenticate users. They are instead part of the internal Linux environment.} Not only does the ability to read this file indicate that the user running the server daemon has superuser privileges; it also makes it possible to attempt a hash-cracking attack on the \textit{root} password set by the manufacturer. The password was hashed using the insecure \textit{MD5 Crypt} algorithm, making it more vulnerable to cracking attacks compared to modern hashing algorithms \cite{provos1999future}. Despite this, it was not feasible to crack the password after an aggressive combination of dictionary and brute-force attacks lasting over a month. This shows that the root password set by the manufacturer is a long and complex string. However, taking control of the device does not require knowing the root password, as this can be achieved by exploiting command-injection vulnerabilities. Furthermore, through the command-injection attack, the root password can be modified to an arbitrary one, hence effectively bypassing the root password authentication. Details of this are discussed later in this paper.

\subsection{Firmware Extraction}

The existence of the file-downloading vulnerability made it possible for the entire file system to be extracted for further examination. By analysing the \texttt{/proc/mounts} file, three files were discovered which, if downloaded, could be used to rebuild the entire file system. This solved the blind file-downloading limitation, as all files could be downloaded at once without having to know (or fuzz) specific paths. Table \ref{tab:mounts} provides further details of these files.

\vspace{-0.8cm}
\begin{table}
\caption{File systems mounted by the device}
\label{tab:mounts}
\begin{tabularx}{\textwidth}{| l | l | X |}
\hline
\textbf{Path} & \textbf{Type} & \textbf{Contents} \\
\hline
\texttt{/dev/mtdblock5} & {jffs2} & Stores user data, such as configuration values. Mounted at \texttt{/etc/jffs2}$^1$. \\ \hline
\texttt{/dev/mtdblock6} & {Squashfs} & Read-only partition for the \texttt{/usr} directory. Stores vendor-specific binaries and scripts, such as startup scripts and the core server application. \\ \hline
\texttt{/dev/root} & {Squashfs} & Stores remaining files that belong in the root folder (\texttt{/}). Includes the Linux kernel and built-in executables. \\ \hline
\hline
\end{tabularx}
\footnotesize{$^1$ In some newer devices, this area is mounted at \texttt{/etc/config}.}
\end{table}

\vspace{-0.4cm}
Having access to the file system made it possible to discover and analyse additional files on the device. This included custom programs such as \texttt{/usr/bin/anyka\_ipc}, the daemon responsible for the UDP service.

\subsection{Remote Code Execution}

\begin{figure}[h]
    \centering
    \includegraphics[width=0.6\linewidth]{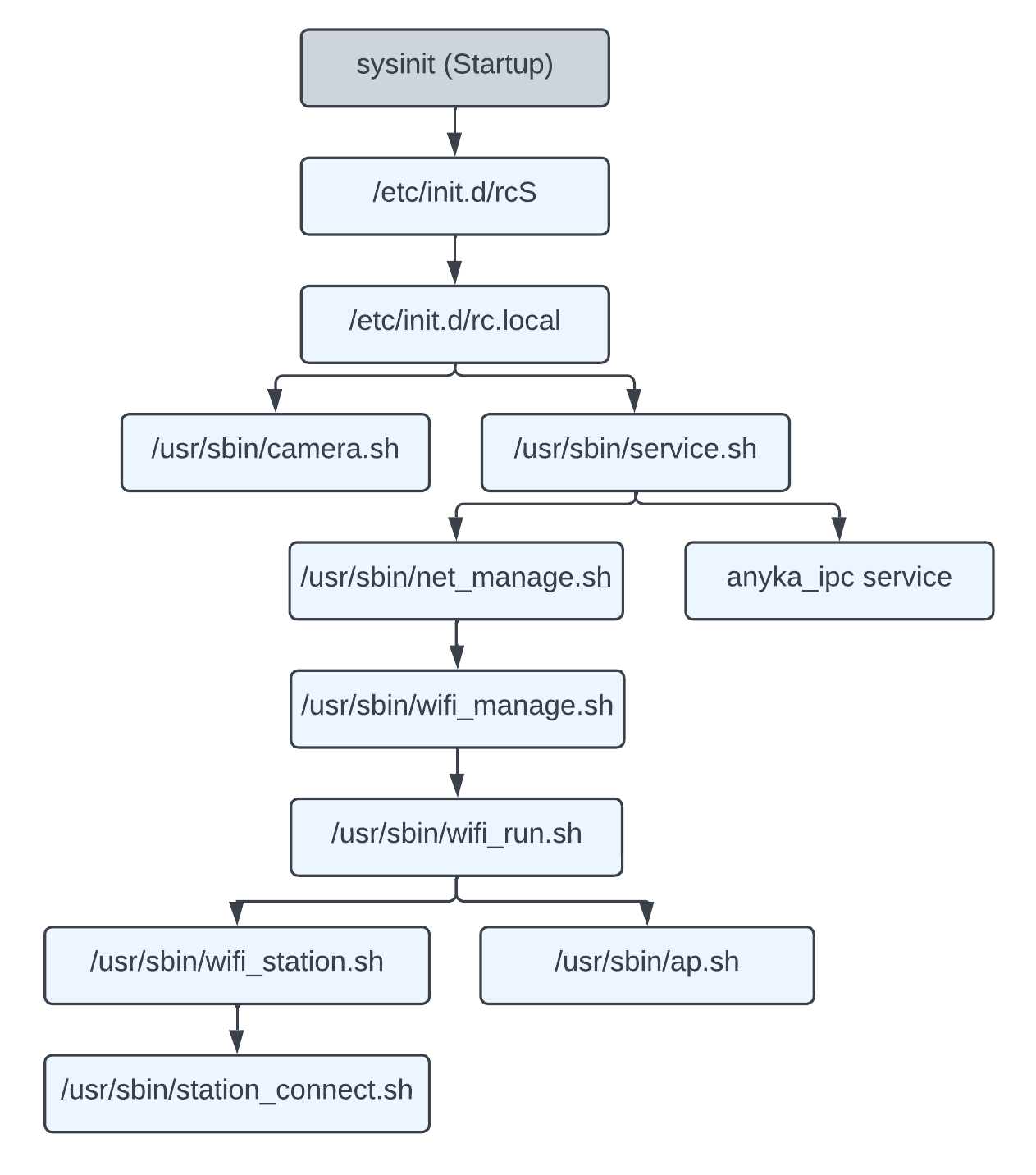}
    \caption{Chain of processes and scripts called on startup}
    \label{fig:startup}
\end{figure}

Analysis of the file system and start-up procedure revealed a chain of bash scripts that are executed on boot, as seen in Figure \ref{fig:startup}. Some of these scripts contain command-injection vulnerabilities that enable an attacker to perform remote code execution on a target device with superuser privileges. 

Vulnerable code has been discovered in modules with software versions as recent as November 2022. A lack of remote updating functionality found in these modules means that it is impossible for patches to be pushed by the manufacturer. An example includes \texttt{station\_connect.sh}, a script responsible for connecting the camera to a user-configured Wi-Fi network. Listing \ref{lst:station_connect} contains a vulnerable excerpt from the script.

\begin{code}
\begin{minted}[breaklines,frame=single]{bash}
SSID=\'\"$GSSID\"\'
PSK=\'\"$GPSK\"\'
...
sh -c "wpa_cli -iwlan0 set_network $NET_ID ssid $SSID"
...
sh -c "wpa_cli -iwlan0 set_network $NET_ID psk $PSK"
\end{minted}
\vspace{-1em}\captionof{listing}{Vulnerable code in \texttt{station\_connect.sh}}\vspace{1em}
\label{lst:station_connect}
\end{code}

The script makes multiple calls to the command `\texttt{sh -c}', which instructs the shell to interpret any following string as a shell command. The variables \texttt{\$GSSID} and \texttt{\$GPSK} originate from the camera's configuration settings (the network name and password), making them directly modifiable by the user, and also an attacker. The danger present is that the user-supplied values are being passed directly into the command, making it possible for a crafted payload to execute arbitrary commands. A weak attempt was made by the manufacturer to prevent this from occurring, however. These values are initially read and parsed in another script, \texttt{wifi\_station.sh}, before being sent to \texttt{station\_connect.sh}. The inputs are weakly sanitised with an \texttt{awk} script (see Listing \ref{lst:sanitise}) that performs the following operations, according to the \texttt{awk} reference \cite{awk}:
\begin{itemize}
    \item Removes all double quotes
    \item Removes leading whitespace
    \item Removes any occurrences of the semicolon (;) character, and any following characters on the same line
\end{itemize}

\begin{code}
\begin{minted}[breaklines,frame=single]{awk}
BEGIN {FS="="}/[wireless]/{a=1} a==1 && 
$1~/^ssid/{
    gsub(/\"/,"",$2);
    gsub(/\;.*/, "", $2);
    gsub(/^[[:blank:]]*/,"",$2);
    print $2
}
\end{minted}
\vspace{-1em}\captionof{listing}{Script to read and sanitise the configuration value for the Wi-Fi SSID}\vspace{1em}
\label{lst:sanitise}
\end{code}

The removal of the semicolon character (and anything after) is a clear attempt to prevent command-chaining. However, not all cases were considered, since additional chaining operators using the ampersand (\texttt{\&\&}) and pipe (\texttt{||}) symbols are never filtered out, which can be used to achieve a similar result. Additionally, the inputs are surrounded with pairs of single and double quotes (see lines 1 and 2 in Listing \ref{lst:station_connect}), in an attempt to ensure the input is interpreted as a string instead of being executed. These techniques, although potentially thwarting a naive command-injection attempt, proved futile since the source code could be viewed. By surrounding the payload with a pair of single quotes, it was possible to break out of the string and achieve code execution.

To perform the attack, an \texttt{OpenWifi} command is sent to the device to update the Wi-Fi settings with the embedded payload. When this is sent, the camera updates its configuration file with the inputs and reboots. On boot, \texttt{station\_connect.sh} is executed, triggering the attacker's code via the call to `\texttt{sh -c}'. A slight barrier to the attack is that the \texttt{OpenWifi} command only supports a maximum length of 32 characters for the SSID and password fields. Recall that the file \texttt{/etc/jffs2/.devpsd} was previously discovered to store the device's password in plaintext. By updating the password to be the contents of a desired script, the password file can be used as a temporary storage mechanism for the payload. The input in the \texttt{OpenWifi} command can then be shortened to execute the contents of this file with the command `\texttt{source /etc/jffs2/.devpsd}'.

Thus, a more sophisticated attack involves sending two commands. The first updates the camera's password to a payload of choice (Listing \ref{lst:updatePwd}), whilst the second updates the Wi-Fi configuration so that the payload is executed on the next boot (Listing \ref{lst:wifiPayload}). It should be noted that no user interaction is required to perform this attack.

\renewcommand{\fcolorbox}[4][]{#4}

\begin{code}
\begin{minted}[breaklines,frame=single]{json}
{
    "cmd": "ModifyPwd",
    "newpwd": $payload, // desired payload
    "pwd": ""
}
\end{minted}
\vspace{-1em}\captionof{listing}{Command to update the device's password}
\label{lst:updatePwd}
\end{code}

\begin{code}
\begin{minted}[breaklines,frame=single]{json}
{
    "cmd": "OpenWifi",
    "sid": $ssid, // user's SSID
    "wifiPwd": "'&&source /etc/jffs2/.devpsd '",
    "state": 1
}
\end{minted}
\vspace{-1em}\captionof{listing}{Payload sent to exploit the command-injection vulnerability in \texttt{station\_connect.sh}}
\vspace{1em}
\label{lst:wifiPayload}
\end{code}

Since the exploit requires an attacker to update the camera’s network configuration, this attack has the side-effect of disconnecting the camera from the Internet, preventing an attacker from sending further commands. This can be resolved by rolling back the credentials after code execution is established. The shell code in Listing \ref{lst:resetWifi} can be added to the payload to restore the original configuration and reconnect to the Internet. Additionally, the password file can be reinstated to its original value to make the attack much harder to detect. By ignoring this step, however, a denial-of-service attack is achieved, since updating the password file is equivalent to changing the password. With the device's password being set to the contents of an arbitrary script, the user will no longer be able to connect to their device via the \textit{LookCam} app.

\begin{code}
\begin{minted}[breaklines,frame=single]{bash}
sed -i 's/^password.*=.*/password = [OLD PASSWORD]/' /etc/jffs2/anyka_cfg.ini
reboot
\end{minted}
\vspace{-1em}\captionof{listing}{Shell code to reinstate the previous Wi-Fi password}\vspace{1em}
\label{lst:resetWifi}
\end{code}

Searching for the vulnerable code segments discovered via \textit{Github Code Search} \cite{githubSearch} and \textit{Sourcegraph} \cite{sourcegraph} revealed that the scripts originate from the 
\textit{AK3918} microcontroller software development kit (SDK). 
Consequently, the command-execution vulnerability is not restricted to the specific modules in this investigation, but potentially to many other products that incorporate the same SDK (or derivatives).

\subsection{Persistent Access}

With the ability to perform code execution, an adversary can perform more sophisticated attacks to persist this access, such as installing a malicious start-up script that exposes a reverse shell. These attacks are immune to the device's `reset' button, as resetting the device only restores the factory configuration file whilst leaving the rest of the filesystem unaffected. In a large-scale attack, this could lead to the formation of a botnet, enabling considerable attacks such as distributed denial of service, botnet mining and mass surveillance. A vulnerable section of code located in \texttt{service.sh} (see Listing \ref{lst:startup.sh}) exposes debug functionality left behind by the manufacturer, making it possible to install a custom start-up script. The code looks for a script located in \texttt{/mnt/usbnet/product\_test}, and if present, executes it on every boot. Additionally, \texttt{Telnet} and \texttt{FTP} (File Transfer Protocol) daemons are started, exposing additional entry points to the camera. 

\begin{code}
\begin{minted}[breaklines,frame=single]{bash}
if test -d /mnt/usbnet ;then # Checks if the directory exists
	FACTORY_TEST=1
...
if [ $FACTORY_TEST = 1 ]; then
    /usr/bin/tcpsvd 0 21 ftpd -w / -t 600 & # Start FTP
    telnetd & # Start Telnet
    echo "start product test."
    /mnt/usbnet/product_test & # Execute the start-up script
...
\end{minted}
\vspace{-1em}\captionof{listing}{Vulnerable debug functionality left behind in \texttt{service.sh}}\vspace{1em}
\label{lst:startup.sh}
\end{code}

An adversary can insert an additional command into the start-up script to change the vendor-set root password to an arbitrary one, as shown in Listing \ref{lst:changeRoot}. This effectively bypasses the root password originally set by the vendor and enables the adversary to authenticate  themselves to the \texttt{Telnet} and \texttt{FTP} services which were previously protected by this root password. 

\begin{code}
\begin{minted}[breaklines,frame=single]{bash}
echo -e "1234\n1234" | passwd root
\end{minted}
\vspace{-1em}\captionof{listing}{Changing the vendor-set root password to `1234' by exploiting the exposed start-up script}\vspace{1em}
\label{lst:changeRoot}
\end{code}

\subsection{Insecure Peer-to-Peer System}

To facilitate remote connections to the cameras outside of the user's local network, the 
peer-to-peer (P2P) system is utilised. Although it offers convenience by enabling users to access their cameras from anywhere in the world, exposing devices to the Internet creates the opportunity for the previously discussed vulnerabilities to be exploited remotely.

\begin{figure}[t]
    \centering
    \includegraphics[width=0.9\linewidth]{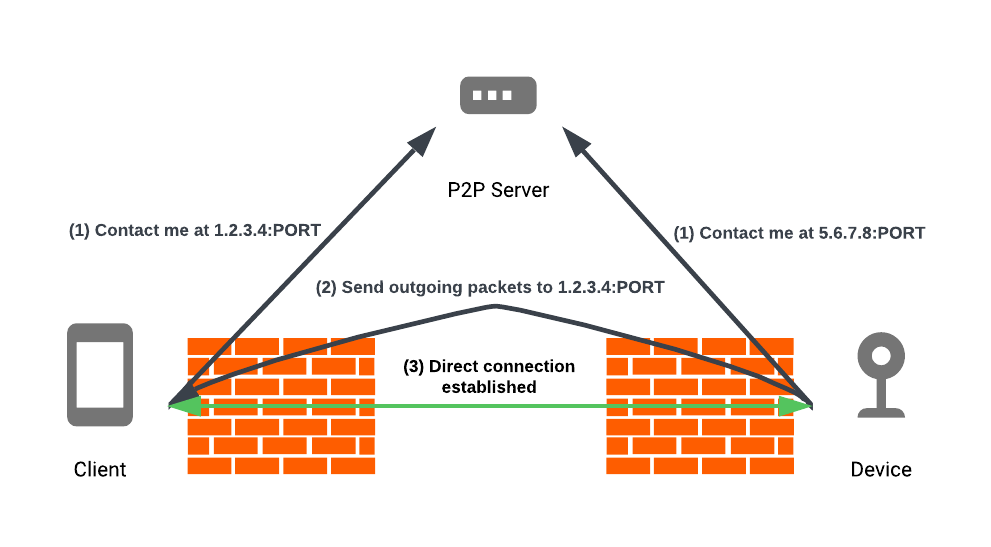}
    \caption{UDP hole punching procedure}
    \label{fig:holePunching}
\end{figure}

The P2P system uses a proprietary security protocol and is inherently insecure. The main role of this P2P network 
is to provide clients with a direct IP connection to the requested device without requiring complex network configuration changes. 
To achieve this, a technique called \textit{UDP hole-punching}~\cite{halkes2011udp} is employed. This method makes it possible for the camera to traverse the NAT (Network Address Translation) system in place within the user's network, essentially performing a port-forward operation without requiring manual changes to the router's settings. It abuses the fact that in many networks, when an outgoing request to a server is made, a temporary NAT rule is created to enable the response to be received. By constantly firing out packets to the client, a `hole' in the NAT table is left open, allowing the app to connect directly to the camera. Figure \ref{fig:holePunching} shows the steps involved in UDP hole punching. The procedure works as follows:
\begin{enumerate}
    \item Both the app and the camera inform centralised peer-to-peer server(s) of their IP addresses and listener ports.
    \item Given that both the phone app and the camera are online, the camera sends outgoing packets to the phone's IP and port to open a NAT hole.
    \item Once the NAT hole is open, the app can connect directly to the camera.
\end{enumerate}

\begin{figure}[t]
    \centering
    \includegraphics[width=6cm]{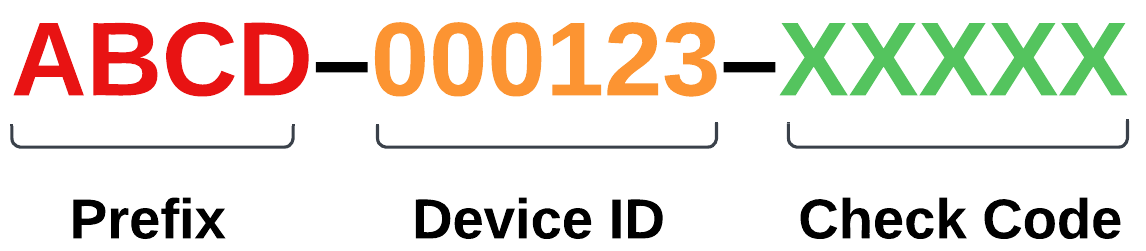}
    \caption{Format of the serial numbers used to identify devices}
    \label{fig:serial}
\end{figure}

Each camera is assigned a unique serial number for identification. When an app wishes to connect to a camera, it sends a request to the peer-to-peer server with the respective serial number. Figure \ref{fig:serial} depicts the serial number format in further detail. Each serial consists of a vendor prefix, an ID number, and a check code (there are 1 million IDs for each prefix; a vendor can license multiple prefixes to support more devices).
The check code is used as an attempt to prevent serial numbers from being enumerated, making it difficult for attackers to guess the serials of other devices. Since serial verification checks have been found to be performed on the server side, it has not been possible to locate the check-code algorithm.

A proprietary encryption method was implemented in the P2P network to protect packets between the apps and  the P2P servers. 
By disassembling the code responsible for this encryption in the \texttt{anyka\_ipc} program using \textit{Ghidra} (a reverse-engineering tool), it was found possible to extract the encryption key and the algorithm used to decrypt packets. 
A function was found within the disassembled program called \texttt{cs2p2p\_P2P\_Proprietary\_Decrypt}, which was reverse-engineered and rewritten in C to decrypt packets captured in \textit{Wireshark}. Multiple keys were located by probing the binary for encryption parameters. Namely, the string 
`\texttt{SSDXXXXXXXXXXXk.}' (part of this string is marked out with `X') is used as a global symmetric key to encrypt and decrypt packets. An additional 256-byte key was discovered that is incorporated as an additional parameter to the encryption/decryption functions. These keys are hard-coded into the binary and are the same for all cameras of the same type. They are also present in the \texttt{libPPCS\_API.so} library included in the mobile app, as it also communicates with the peer-to-peer system.

With the ability to communicate with the P2P servers, an attacker can 
request the IP addresses of cameras, enabling a direct connection to be made. It is important to note that this vulnerability does not only apply to the cameras in question, but to any IoT device using this network to facilitate P2P connections. This makes it possible for vulnerabilities present in the spy cameras and also other IoT devices to be exploited remotely, since no authentication is required to gain a direct IP connection. This raises concerns surrounding potentially many more products in the IoT space.

To sum up, these generic hidden camera modules have exhibited considerable defects in their various components, involving multiple companies in the supply chain. The network service running on each camera to support the mobile apps contains several vulnerabilities that enable attackers to bypass the authentication system and extract sensitive information. The insecure configuration scripts included as part of the microcontroller SDK make the cameras vulnerable to command-injection attacks. Poor system configuration enables the command injection attacks to be performed with superuser privileges. A flawed encryption system in use by the peer-to-peer system enables attackers to impersonate legitimate users, exposing IoT devices to the Internet and allowing attacks to be performed on an international scale, to potentially many millions of devices.

\section{Mitigation measures}

To protect these cameras, a complete overhaul of the system would be necessary. This is due to the numerous vulnerabilities present in all of their components. Despite several attempts to bring these issues to the attention of the responsible parties, many refused to respond or cooperate whilst others expressed no interest. The lack of success in identifying remote updating functionality in the system suggests that any mitigation attempt through patching would be infeasible nevertheless. 
A list of remedial actions has been compiled below to address each of the identified vulnerabilities for the future development of IP cameras:

\begin{enumerate}
    \item \textbf{Verbose debug logging} -- Strip out debug messages in the production build of the controlling app.
    \item \textbf{Unencrypted communications} -- Implement Transport Layer Security (TLS) to protect traffic in-transit.
    \item \textbf{Authentication loophole} -- Issue a token on successful login and verify the presence of a valid token in the subsequent requests.
    \item \textbf{Unencrypted password storage} -- Store a salted hash of the user's password instead of storing it in plaintext.
    \item \textbf{Arbitrary file downloading} -- Associate each video file with an identifier and have an internal database mapping identifiers to file paths. With this method, a user only inputs an identifier instead of a path. Thus, the input is not directly incorporated into the path, avoiding directory traversal.
    \item \textbf{Unnecessary exposure of sensitive information} -- Avoid printing sensitive information such as Wi-Fi credentials in command responses, e.g., \texttt{GetDevInfo}.
    \item \textbf{Poor access control} -- Create a non-privileged Linux user to run the application server. Modify file/directory permissions to protect privileged information.
    \item \textbf{Outdated password hashing algorithm} -- Ensure password hashes for all users are using a modern algorithm such as \textit{bcrypt} \cite{provos1999future}.
    \item \textbf{Command injection} -- Modify shell scripts to correctly sanitise user input. Alternatively, rewrite the functionality in another language such as C to mitigate the risk of command injection.
    \item \textbf{Flawed encryption in P2P network} -- Implement a TLS layer to protect traffic in-transit.
    \item \textbf{Device impersonation} -- Adopt a secure enrolment process to register a camera  with the app, e.g., based on Thread~\cite{thread}, to create end-to-end secure channels between the camera and the controlling app without having to trust any peer-to-peer servers. 
    
\end{enumerate}

\section{Future Work}

It has been demonstrated how attackers can perform remote code execution on an arbitrary spy camera with only the knowledge of its serial number. The proprietary check-code algorithm used to verify serial numbers serves as the sole defence against device enumeration, which could lead to the formation of a botnet of potentially millions of vulnerable devices. Serial enumeration does not only impact the hidden cameras investigated in this paper, but also any type of IoT device connected to the P2P network. Recalling that over 50 million IoT devices are estimated to be using this system, the possible impact here can be much greater. Flaws in the design of the network and its encryption protocol make it possible to gain a direct IP connection to arbitrary devices without supplying any credentials. Further exploits could be possible based on the designs of these devices. A possible route to cracking the check-code algorithm would be to purchase a copy of the P2P server software from the manufacturer (which costs around \$1,000), so one has access to the source code of the server software including the implementation of the check-code algorithm. With the knowledge of the check-code algorithm, an attacker may extend the reported attacks to an arbitrary IoT device with a valid device ID in the P2P network, including not only IP cameras but also IP-based smart locks, doorbells, bulbs, light switches, speakers and so on. We leave this to future study.

\section{Conclusion}

A systematic investigation of the security of IP-based hidden cameras has been conducted, revealing a broad range of vulnerabilities. These vulnerabilities allow a remote attacker, with the mere knowledge of the camera's serial number, to take complete control of the camera even if the camera is within an internal network behind a firewall. Proof-of-concept attacks have been demonstrated to eavesdrop on the audio and video streams, retrieve any recorded video stored on the camera module along with other sensitive information (such as the Wi-Fi passwords of the user's home network), and run a reverse shell script on the camera device (by abusing the password update function and specifying the reverse shell as part of the input to that function), thus turning the camera into a platform to attack other nodes in the home network or as part of a botnet. These attacks are not just limited to hidden cameras; they are generally applicable to IoT devices that follow a similar security design. Countermeasures are proposed to contain these attacks. However, patching or recalling the affected cameras is infeasible given the existing designs of these products and the complexities of the supply chain. Manufacturers are urged to pay more attention to security and get it right at the start, as failures can cause unintended, severe, and long-lasting consequences, especially when retrospective fixes are impossible. In the meantime, the public should be informed of the security issues of a hidden camera, especially about the danger of sharing a camera's serial number with others. Even if a user diligently does not share the serial number, we caution that an attacker may already know it, e.g., by enumeration, or reading the product information in the supply chain. To ultimately address the vulnerabilities identified in this paper, we call for open, peer-reviewed and standardised security designs, which are currently lacking for hidden cameras and similar IoT products.   

\subsection*{Acknowledgements} The second author is supported by
Royal Society (ICA\textbackslash R1\textbackslash 180226) and EPSRC (EP/T014784/1).

\bibliographystyle{splncs04}
\bibliography{bibliography}

\end{document}